\documentclass[10pt,a4paper,twocolumn]{article}

\date{}

\author{H. GENREITH
\footnote{\em Email:genreith@geo.uni-koeln.de,
WWW:http://www.uni-koeln.de/math-nat-fak/geomet/geo/mitarb/genr.html}
\\Institut f\"ur Geophysik und Meteorologie 
der Universit\"at zu K\"oln}

\title{A Black Hole emerged Universe}

\begin{document}

\maketitle

\begin{abstract}
A simple model of a Universe
is presented
\footnote{Contribution-desk paper on the $8^{\rm{th}}$
International Conference on the
Structure of Baryons, BARYONS'98, Bonn, Sept. 22-26, 1998.} 
composed of black holes
and black branes. It uses the most simplest approximations and models of
General Relativity and Quantum Dynamics to offer an idea of an unification
and gives a possible answer to the quantization and entropy of Black Holes.
It proposes a mass spectra for elementary particles and gives a vivid
interpretation of the particle-wave-dualism.
\end{abstract}

\section{Introduction}
A simple model for the construction of fermionic matter by
socalled black-holes will be suggested.
This paper gives a short overview on cosmology in its first part. 
In its second part it proposes 
a possibly closed model for fermionic matter from neutrinos
to the Universe itself,
based on the concept of black-holes/black-branes.

\subsection{The 'embedding' problem}

The recent discussion on the origin and destiny of the Universe
raises the question
where the Universe comes from. The main explanation is that time and 
space were created with the big-bang and therefore the
question, where the Universe is embedded in or what happened before the
creation, is not permissible~\cite{hawking}
\footnote{The nonpermissibility of the question follows
from the fact, that Riemannian Geometry
can describe the metric intrinsic, so an outer space is
mathematically not demanded. But it also
doesn't exclude the possibility.}. This implies the assumption
that there was no reason for the creation and
that the Universe was created
from 'nothing' by a kind of quantum fluctuation and should have a
zero overall-energy. Although this explanation sounds well, it doesn't
solve the problem really: If the Universe was created from 'nothing',
why shouldn't be there uncountable more Universes; and what relationship
could there be between them ? Numerous scientists tried to
give an explanation
to this question (e.g. A. Linde~\cite{linde}).
 Whatever the correct answer to this
question is, the explanation must be a kind of an infinite regress as
otherwise the embedding problem would recur.

\subsection{The 'matter-grain' problem}

A still unsatisfactory answered question is, what matter, like
elementary particles, really is.
To answer this question, one tries to find the most elementary particles
by experiment and theory.  All matter should be composed of  this
smallest 'grains' of matter. But this explanation
always raises the question
that these 'grains' could be composed of something,
that can be described by
a more elementary description. Here comes in the socalled string-theory,
which
gives a topologic explanation, the strings, as the origin of matter.
But what
are these strings composed of ? They should be pure topology,
like space-time
curvatures are~\cite{hooft,kaku}.

\subsection{The mass of the known Universe}

The mass\footnote{The 'mass' of the Universe can be defined only if 
there is a border and an corresponding volume. So the Universe mass can be
defined in the volume defined by the world radius (which is defined by an 
infinte redshift) or by the (hypothetical) Schwarzschild-radius.}
of the bright
matter lying in the reach of modern observation techniques may be
derived for
example from the outcome of the Hubble deep field: On this image made 
by the Hubble space telescope, which
contains galaxies down to about 29 th. magnitude, one
can calculate about 1 Million Galaxies on an area of 1 square degree.
With usual masses for galaxies this means that the mass of the bright
matter reaches an amount of approximately
some $10^{51} $ kg~(Sexl)~\cite{sexl}.
The critical mass density of the Universe is the amount of
mass which is needed to
bring the universal expansion to a halt in the future  and to a collapse
in a final big crunch. This mass is, depending on the world model,
about $10^{53} $ kg. This means that the visible mass of the Universe
is sufficient for only about some percent of the critical mass density.

On the other hand the mass density of the baryons derived from the theory 
of nucleosynthesis and the measured photon density should be 10 to 12
percent of the critical mass density~\cite{coughlan}.
Therefore the so-called dark matter should be about 10 times
greater than the visible bright mass. This matter shows itself in the
extinction of light on its way through the Universe as well as by its
gravitational force on galaxies and galaxy clusters~\cite{durrer}.

Nevertheless the mass density of baryonic matter is with that
only about a tenth of the mass
density needed to close the Universe. Besides of the classical
baryonic matter therefore also
exotic matter gets into 
considerations on the overall mass of the Universe. These are,
for example, a not vanishing rest
mass of the 
neutrino, super-symetric elementary particles as e.g. gravitinos,
extremely weak interfering
particles
(WIMPS), and black holes of various sizes.
So there exist a lot of estimations of the Universe mass density
up to several times
the critical mass density as is
$\rho_{\rm{U}} \cong 0.01 - 2.0\cdot\rho_{\rm{critical}}$ in
Coughlan~\cite{coughlan} or
$\rho_{\rm{U}} \cong 0.01 - 10.0\cdot\rho_{\rm{critical}}$ in
Kaku~\cite{kaku}.

\subsection{The Schwarzschild-metric and the primordial
expansion of the Universe} \label{SSM}

The Schwarz\-schild solution $R_{\rm{ss}} = \frac{2GM}{c^2}$ for the
Schwarz\-schild-radius of a given mass $M$ is derived from the
Einstein field equations
considering a point-like mass. The Schwarzschild solution is a static,
homogenous and isotropic
solution for the region outside the Schwarzschild
radius ('A black hole has no hairs').
The inside solution may have other solutions, the most interesting is the
solution of Oppenheimer and Snyder~\cite{sexl,oppenheimer} which shows the
astonishing result, that the inside solution must be a Friedmann-Universe.
This results from the fitting of the outside to the inside solution of a
collapsing star: After the burning out of a star every pressure
vanishes when
the star shrinks to 9/8-th of its Schwarzschild-radius.
When the collapsing
star reaches the event-horizon one has to set $p=0$ and one gets
a Friedmann-Universe
at its maximum expansion. So the Oppenheimer/Snyder-solution maybe
taken for a
speculation of a Friedmann-Universe inside a black hole, shrinking
down to a
singularity and from there expanding back to its
maximum expansion again.

If one don't like such an interpretation, there is the problem
to explain how the expanding
Universe let behind its own event-horizon.
As the Universe expands starting at vanishing dimensions,
this leads to the a paradox
looking context:
Either the Schwarz\-schild-radius is greater-equal to the
world radius\footnote{Although the
dimension of the world radius and the Schwarzschild-radius may
be equal, there is a considerable
difference: The world radius is the 'visible' dimension which is
defined by an infinite red
shift. 
On the other hand, the  Schwarzschild-radius can never be seen even
by an observer
placed close to this border as seen from 'outside'. In this case the
line of sight would be
curved when seen from outside but an inside observer would observe
it as free of any forces and
straight. By this any number of world radius' can be placed into the
area of the Schwarzschild-
radius even if both have the same dimension.} of the Universe, as
large estimations of the
Universe mass assume, then the Universe may be called a black hole or
white hole, because of the
time-reversal
of the dynamics.
Or, as small estimates of the mass of the Universe assume,
the Schwarz\-schild-radius is  about
10 percent of the today world radius of the Universe.
Now then the Universe should have expanded
beyond\footnote{A 'pushing in front' of the event-
horizon is easilly possible due to the increasing scale-factor $R(t)$
of the Universe.}
its Schwarz\-schild-radius in former times after the big bang when it
crossed a radius of about
1.5 billion light-years which is an event that is supposed to
be not in harmony with general
relativity.

By this one could guess that our Universe should be indeed a
black hole\footnote{A black hole, if watched from inside,
may be called a white hole because it is guessed that from its inside
singularity everything can come into existence~\cite{kiefer}.
A black hole actually needs an outer space which it is defined on,
but the existence or not of an
outer space is first of all a question of belief.}.
The theory of inflation~\cite{linde} tries to avoid such a paradox by a
kind of extremely fast inflationary expansion.
But the inflation phase ends when the Universe is some cm in size, much
smaller
than its least possible event-radius. And last the inflationary
model also demands a critical
mass density for the Universe.

But even if one believes in a Universe which is not of critical mass
density in its visible worldradius and which should expand over this
range up
to infinity, there will always be a radius where the mass density
of the inner region gets critical:
as vacuum is never empty of energy and as the Schwarzschildradius
grows linearly
with the mass, while the mass of a volume grows with the 3rd power of the
radius:
$ \rho_{\rm{c}}=3c^2 / 8 \pi G R^2
\Leftrightarrow   R_{\rm{c}}=\sqrt{3c^2 / 8 \pi G \rho} $
This gives a critical value for the mass density of the Universe of some
$10^{-30}\frac{\rm{g}}{\rm{cm^3}}$ if one considers the area of the visible
worldradius which is $R=2c/H_0$ in the socalled EdS-model.
So if the mass density would be greater-equal to this
value, our Universe may
be explained as a black hole; if the value is less,
one could avoid such an
explanation if one claims that the regions outside the related
event-horizon are gravitational incoherent. But the high uniformity of
the 3K-background radiation and also the theory of inflation indicate
that they are indeed coherent.

A lot of numerical simulatuions of Universes with different parameters
were done \cite{frenk,vaas} (and will go on). A lot of models prefer
cosmologies with
$\Lambda \ne 0$ and small densities, which are also in agreement with the
interpretation of the luminosities of far away Type-Ia-Supernovae,
which seem to be about $0.3m$ darker than expected. But the recent work of
Gawiser and Silk
\cite{gawiser}, which numerically calculates the 10 most
discussed cosmological models and
relates them to the variation of the observed cosmic microwave background
(CMB), shows the astonishing result,
{\em that the critical standard model fits the observation of the CMB
by far best}.

In this sense a black-hole-Universe is defined through a Universe of
critical mass density. The possibility of a black-hole-Universe was already
taken into consideration e.g. in J-P.Luminets
~\cite{luminet} book 'Black Holes' or in a controversal darwinistic
view in the recent work of Smolin~\cite{smolin,concar}.

\section {Simple solutions for a black-hole-Universe}

\subsection{The Einstein-deSitter model}

The Einstein-deSitter model (EdS model) follows from the Friedmann-model
simplified with $\kappa=0$ and $\Lambda=0$ which means an euclidic,
isotropic and
homogenous world model.  Then the scale-factor  $R(t)$  is a solution of
$ R \dot R^2 = \textrm{Const.} $ which gives
\begin{equation}
R(t_{\rm e}) = R_{\rm 0}\cdot { (\frac{t_{\rm e}}{t_{\rm 0}}) }^{2/3}
\label{EdSSkala}
\end{equation}
in which $t_{\rm{e}}$ is the time of emission of a signal 
( $t_{\rm{e}}=0$ is the time of the big bang) and 
$t_{\rm 0}$ the time today. 
The relationship for the age of the world follows from the Hubble equation
\begin{displaymath}\frac{dR}{dt} = H(t) R(t)
\quad \textrm{which gives} \qquad	t_{\rm 0} = \frac {2} {3H_{\rm 0}}
\end{displaymath}
The actual distance  $r$ between two points with distance $\rho$
is derived from the equation:
\begin{displaymath} r(t) = R(t) \cdot \rho \end{displaymath}
The standardization is given referring to the present time by
$ R(t_{\rm 0}) = R_{\rm 0} =: 1 $~.

General Relativity demands a maximal velocity  $c$ only for the
peculiar movement.
The variation of the scale-factor
\begin{displaymath}
\frac{dR}{dt_{\rm e}}
= \frac{2 R_{\rm 0}}{3  t_{\rm 0}^{2/3} t_{\rm e}^{1/3}}
\end{displaymath}
runs to infinity for small times of emission.
Therefore the overall-velocity of the Expansion
\begin{equation}
\frac {dr} {dt} = \frac {dR} {dt}
\cdot \rho + R \cdot \frac { d \rho } {dt}
\label{drdt}
\end{equation}
can be much greater than the speed of light.
Considering events for which the speed of light is a given limit
one has to look upon variations
of  $\rho$.

The red shift  $z$ is interpreted as the scale variation of the
wavelength of a photon:
\begin{equation}
\frac {\lambda_{\rm 0}} {\lambda_{\rm e}}
= \frac{ R_{\rm 0}} {R_{\rm e}} =: 1 + z
\end{equation}
The fact that the EdS-model is simplified with a curvature
parameter $\kappa=0$
seemes to point out (as (\ref{drdt}) never turns its sign to minus)
that it is not suitable to
treat a black-hole-Universe which should have an $\kappa=+1$
and should collapse in a finite
time. But in the scope of this simple model  it means that a
critical EdS-Universe is a black-
hole-Universe collapsing in an infinite time.

A derivation of the EdS model can be gathered from the current
literature, e.g. Sexl~\cite{sexl}, Schulz~\cite{schulz},
B\"or\-ner~\cite{borner}.
Because of its considerable simplifications the EdS model is
analytical treatable and therefore
is an excellent tool for the theoretical handling of cosmological
phenomena.
On the other hand this model doesn't fit for instance for large
values of the Hubble constant $H_{\rm 0}$ which leads to a smaller
age of the world than observational results suggest. So recently world
models are brought into discussion which are determined
by a  $\Lambda\neq 0$~\cite{vaas,priester,frenk}.
These models are better compatible with large values of the Hubble
constant; they are mathemathically more complex and
have to be calculated partly numerically.

\subsection{The Planck dimensions} \label{diePldims}

If one equates the Planck energy $E = \hbar\omega $
with the  Einstein energy $ E = m_{\rm 0}c^2$ for the rest energy
of a mass charged particle
one gets the de-Broglie wavelength of a resting particle $m_{\rm 0}$.
This is also known from the theory of photon scattering on electrons
as the Compton wavelength
$\lambda_{\rm C}=\frac{\hbar}{m_{\rm 0}c}$.
This wavelength is opposite proportional to the mass as the
Schwarz\-schild-radius
$\rho_{\rm{ss}}=\frac{2Gm}{c^2}$.
The identity of both lengths leads to the Planck dimensions
\footnote{The usual description is $m_{\rm{pl}}
=  \sqrt{ \frac{c\hbar}{G} }$ from just
dimensional arguments. I will use the definition following
from the equality of event-radius to
Compton-wavelength. See also footnote following equation (\ref{spin})}:
\begin{eqnarray}
m_{\rm{pl}}  & = &  \sqrt{ \frac{c\hbar}{2G} }
=	1.54\cdot10^{-8} \rm{kg}	\nonumber\\
l_{\rm{pl}}  & = &	\frac{  \hbar  }{ m_{\rm{pl}}  c }
=	\sqrt{ \frac{2 \hbar G }{c^3} }
=	2.29\cdot10^{-35} \rm{m}	\nonumber\\
t_{\rm{pl}}  & = &	\frac{ \hbar }{  m_{\rm{pl}}c^2  }
=	\sqrt{ \frac{2 \hbar G }{c^5}  }
=	0.762\cdot10^{-43} \rm{sec}  \nonumber\\
E_{\rm{pl}}  & = &  m_{\rm{pl}}c^2
=  \sqrt{ \frac{c^5\hbar}{2G} }
= 1.38\cdot10^{9}  \rm{J}
%%%=  0.864\cdot10^{28} \rm{eV}
\label{Pldim} \end{eqnarray}
The main property of the Planck dimension is the fact that
the energy of a wave with wavelength
$l_{\rm{pl}}$ equals to a mass which bends space  to a black hole of
Planck size.

\subsection{The mass formula for the Einstein-de Sitter model}

The Eds model delivers the coordinate distance of an event
travelling with the speed of light
from (\ref{drdt}) through the integration of
$ d\rho = \frac {c \cdot dt} {R(t)} $:
\begin{equation}
\rho=\frac{3ct_{\rm 0}}{R_{\rm 0}}\cdot
\left[ 1-  (\frac{t_{\rm e}}{t_{\rm 0}})^{1/3}
\right]
\label{pkentf}
\end{equation}
This distance should be equal at maximum to any given Schwarz\-schild-radius
$\rho_{\rm{SS}}=2GM/c^2$ which results in:
\begin{equation}
M(t_{\rm e})= \frac{c^3} {H_{\rm 0}G} \cdot
\left[  1 - \left(\frac{3 H_{\rm 0}t_{\rm e}}{2}\right)^{\frac{1}{3}}
\right]
\label{MF}
\end{equation}
This time-dependent mass is the mass to be at least included by
a gravitational spherewave
starting at time $t_{\rm e}$ and running with velocity $c$.
Otherwise the wave would have to go
beyond  its own event-horizon.
Inserting $t_{\rm e} = 0$ for the origin of the Universe one gets:
\begin{equation}
M_{\rm{Umin}} \ge M(0)
= \frac{c^3}{H_{\rm 0}G}
\label{MU}
\end{equation}
This is the mass which makes an EdS Universe critical.
The value of $H_{\rm 0}$ is still controversal and differs
depending on the source between
approximately $50$ and $100 \frac{\rm{km}}{\rm{sec} \rm{Mpc}}$.
So the mass of the Universe should be in the range of $M_{\rm U}
\in  [1.248,2.497] \cdot10^{53}
\rm{kg}$ .

Since the geometry of space-time in our Universe yet depends only
on the mass included in the
Schwarz\-schild-radius, the equality sign in (\ref{MU}) should be right:
The mass of the Universe gets herewith the rank of a constant of
nature as it is expected for a
black hole. In this case one may formulate:
\begin{equation}
c  =\sqrt[3]{M_{\rm U}H_{\rm 0}G}=:\alpha\sqrt[3]{M_{\rm U}}
\nonumber \\
\label{cMU}
\end{equation}
which means that a black-hole-Universe would relate the speed
of light to the mass and expansion
rate of the Universe.
One may define a topologic constant  $\tau$  for the EdS model as
\begin{equation}
\tau :=M_{\rm U}H_{\rm 0}=\frac{c^3}{G}
=4.038\cdot10^{35}\frac{\rm{kg}}{\rm{sec}}
\label{tEdS}
\end{equation}
which is the product of the Universe expansion rate and mass and
relates $c$ and $G$ as
$c=\sqrt[3]{\tau \cdot G}$.

\subsection{The resulting force on elementary level}

One may generalize the mass formula (\ref{MF}) to local
gravitational waves running in a local
flat space-time. For that purpose (\ref{MF}) is expanded to a
Taylor series at the time
$t_{\rm 0}$ transforming the time coordinate to $t=t_{\rm 0}-t_{\rm e}$. 
By this one gets the mass formula for small masses implieing
small times $t<<t_{\rm{0}}$:
\begin{eqnarray}
m(t) =  \frac{1}{2} \tau t
+ \frac{1}{6} \tau  \frac{t}{t_{\rm 0}} \cdot t
+ \frac{5}{54} \left( \frac{t}{t_{\rm 0}} \right)^2 \cdot t
+ R(O^4)
\nonumber\\
 =  \frac{1}{2} \tau t
+ \frac{1}{4} \tau H_{\rm 0} \cdot t^2
+ \frac{5}{24} \tau H_{\rm 0}^2 \cdot t^3
+ R(O^4)
\label{kleinmt}
\end{eqnarray}
As the factor
$ \left(\frac{t}{t_{\rm 0}}\right)^{n}$
rapidly drops to zero for small times one can calculate
further on with only the first part of
the sum:
\begin{equation}
m(t) = \frac{1}{2}\tau t
= \frac{1}{2}\frac{c^3}{G}\cdot t
\label{kleinm}
\end{equation}
Equivalent to the derivation of  (\ref{kleinmt}) one gets for
the peculiar distance of events
travelling with the speed of light considering small times:
\begin{equation}
\rho (t) =  c t  +  \frac{c H_{\rm 0}}{2} t^2
+  \frac{5}{12} c H_{\rm 0}^2 t^3  + R(O^4)
\end{equation}
From there the apparent force\footnote{This force is seen only
by a observer from far away. An
observer travelling inside a black hole would never feel to
hit against the event-horizon because any forced curvature of his
line of sight would be sensed
to be straight.} acting on a gravitational event running with $c$ is:
\begin{equation}
\mid \vec F_{\rm{e}} \mid
= \frac{d}{dt}(mv)
= \frac{c^4}{2G} (1 + 3H_{\rm 0} t + R(O^2)) \cong \frac{c^4}{2G}
\label{Fe}
\end{equation}
This practical constant force acts on the event over the area
of the Schwarz\-schild-radius:
\begin{equation}
E \approx F_{\rm{e}}\cdot \rho_{\rm{ss}}
= \frac{c^4}{2G} \cdot \frac{2Gm}{c^2}  =  m c^2
\label{Emc}
\end{equation}
As it seems a gravitational wave may run unhindered just if
the mass included in its sphere is
zero or infinite.
Every distortion\footnote{A distortion of such a kind is given in
general by the deviation of the
space-time-curvature through the gravitational wave itself if the
energy of the wave is not
neglectible.}
of space-time causing a mass creates an event-horizon proportional
to this mass. After that the
wave hits its selfmade horizon. The rest energy of the so
created (virtual) mass is borrowed from
the energy of the gravitational wave and is transformed into an
potential energy of  the same
quantity represented by a virtual
black hole\footnote{A 'virtual' black hole differs from a
'real' black hole by the fact that the energy for its creation
does not come from a fatal
gravitational collapse and by this leaving behind a large potential
well but was borrowed from
the energy of a gravitational wave.}.

As a result of S. Hawking the power of radiation~\cite{sexl} of a
black hole is approximately $P \approx \frac{\hbar c^6}{G^2M^2}$.
As the energy is $E = P\cdot t \cong Mc^2$ this results in an
approximately time for vaporizing of
a black hole $t_{\rm V} \cong \frac{G^2M^3}{\hbar c^4}$.
Therefore one can make the approximation for the evaporating time
for a black hole of Planck mass:
\begin{equation}
t_{\rm V} \approx \frac{G^2}{\hbar c^4}
\left(\frac{c\hbar}{2G}\right)^{\frac{3}{2}}
= \frac{1}{4} \sqrt{\frac{2 \hbar G}{c^5} } \approx t_{\rm{pl}}
\label{tv}
\end{equation}

In this model primordial space-time is a highly rigid medium
to the ('primary') gravitational
wave which meets its motion with a titanic force
$F_{\rm{e}}  = \frac{c^4}{2G}\cong 6\cdot10^{43}
N$.
The wave is broken down to small fluctuating areas of space-time
riding like foam on
ocean waves~\cite{hawking,hooft,luminet}. Although the primary wave
is stopped by its self-
curvature of space,
its overall-velocity (the 'secondary' gravitational wave)
still is in the
order of the speed of light as an area of Planck's dimension
evaporates in a time equal to
Planck's time:
\begin{equation}
v \approx \frac{l_{\rm{pl}}}{t_{\rm{pl}}}  = c
\label{vgw}
\end{equation}

\section{An approximate stationary solution for black hole particles}
\label{stNSL}

Because the elementary force  $\vec{F}_{\rm{e}}$ always
obstructs the movement of the
gravitational wave the integral of energy on a closed loop is not zero.
The constant force $\vec{F}_{\rm{e}} = - \frac{c^4}{2G} \cdot \vec{\rm e}$
has just a pseudo-potential
\begin{equation}
V( r )  = \frac{c^4}{2G}\cdot r
\label{Pot}
\end{equation}
As a simple approximation one can consider the effect of
a wave in a rectangular potential well.
The gravitational wave runs unhindered in a small area until
it is stopped, as mentioned from
outside, by its selfmade event-horizon which exerts a nearly
infinite apparent force of
$F_{\rm{e}}$ hindering the wave\footnote{One may also imagine
this wave as a wave travelling
around the
event-horizon with $c$.} in going on:
\begin{eqnarray}
V & = &0  \qquad  \,\, \textrm{for} \quad r \in [0,\rho_{\rm{SS}}]
\nopagebreak[3]
\nonumber\\
V & = & \infty \qquad  \textrm{for} \quad r >  \rho_{\rm{SS}}
\label{Pot0}
\end{eqnarray}
The common known solution of the time-in\-de\-pen\-dent
Schr\"o\-din\-ger equa\-tion for a
particle in such a rectangular potential well delivers
the energy eigenvalues
\begin{equation}
E_{\rm{n}}=\frac{n^2\pi^2 \hbar^2}{2ma^2}
\end{equation}
with $n=1,2,3\dots$ and  $a$ the diameter of the well.
With the substitution of $a = 2\rho_{\rm{SS}}= \frac{4Gm}{c^2}$  and
$m = \frac{E_{\rm{n}}}{c^2}$
one gets the  energy eigenvalues as the real roots of
$ E_{\rm{n}}^4=\frac{n^2\pi^2 \hbar^2 c^{10}}{32 G^2} $~:
\begin{equation}
E_{\rm{n}} = \pm \frac{\sqrt{\pi}}{2^{3/4}}  \sqrt{n} E_{\rm{pl}}
\end{equation}

The factor $ \frac{\sqrt{\pi}}{2^{3/4}}  = 1.054$ origins
from the simplification of the
potential (\ref{Pot0}) and is set equal to 1.
Then the masses of the virtual Schwarz\-schild areas have the
eigenvalues
\footnote{The same quantization rule for Black Holes was found
by Khriplovich~\cite{khriplovich}(1998) from a totally different
point of view.}
\footnote{For the handling of negative masses see also the article
of Olavo\cite{olavo}}:
\begin{equation}
m^{\pm}_{\rm n}=\pm \sqrt{n}\cdot m_{\rm{pl}}
\label{mn}
\end{equation}
The difference of neighbouring positive eigenvalues is with this
\begin{equation}
\Delta m_{\rm n} = m_{\rm{n+1}} - m_{\rm n}
= ( \sqrt{n+1} - \sqrt{n} ) m_{\rm{pl}}
\end{equation}
which can be written for large  $n\rightarrow \infty$ :
\begin{equation}
\Delta m_{\rm n} = \frac{1}{2\sqrt{n}}\cdot m_{\rm{pl}}
\label{dmn}
\end{equation}
From (\ref{mn}) and (\ref{dmn}) follows
\begin{equation}
\Delta m_{\rm n} = \frac{1}{2} \cdot \frac{m_{\rm{pl}}^2}{m_{\rm n}}
\qquad \textrm{with} \qquad
n= \frac{1}{4} \cdot \frac{m_{\rm{pl}}^2}{\Delta m_{\rm n}^2}
\label{dmn1}
\end{equation}
and the relation between stimulated mass $\Delta m_{\rm n}$ 
and  positive virtual mass $m_{\rm n}$ is:
\begin{equation}
\frac{\Delta m_{\rm n}}{m_{\rm n}} =  \frac{1}{2n}
\end{equation}
The radius of an elementary particle in this model is the
event-radius of the virtual mass
\begin{equation}
\rho_{\rm n} := \rho_{\rm{SS}}(m_{\rm n}) = \frac{2Gm_{\rm n}}{c^2}
\end{equation}

The formula (\ref{dmn1}) therefore serves the equation
\begin{displaymath}
\lambda_{\rm C} = \frac{\hbar}{m_{\rm 0}c}
\qquad \Leftrightarrow \qquad
m_{\rm 0}\cdot \lambda_{\rm C} = \frac{\hbar}{c}
\end{displaymath}
for the Compton wavelength of an elementary particle:
\begin{equation}
\Delta m_{\rm n} \cdot 2 \rho_{\rm n}
=  \frac{m_{\rm{pl}}^2}{2m_{\rm n}} \cdot \frac{4Gm_{\rm n}}{c^2}
=  \frac{\hbar}{c}
\end{equation}
with the substitutions $m_{\rm 0}=\Delta m_{\rm n}$ and $\lambda_{\rm C}
= 2\rho_{\rm n}$.

During the creation of the virtual black hole Heisenberg's
uncertainty relation is fulfilled for
the stimulated mass:
\begin{equation}
\Delta E \Delta t
= \Delta E_{\rm{n}} \cdot \frac{\Delta x}{c}
= \Delta m_{\rm n}c^2 \cdot \frac{2Gm_{\rm n}}{c^2\cdot c}
= \frac{Gm_{\rm{pl}}^2 }{c} = \frac{\hbar}{2}
\end{equation}

The angular momentum $ \vec{L}=m\vec{v} \times \vec{r} $ of a
wave having mass $\Delta m_{\rm n}$
circulating on the horizon of a black hole with mass $m_{\rm n}$ at
the speed of light $c$ is:
\begin{eqnarray}
\mid \vec{L} \mid & = &\Delta m_{\rm n} c \cdot \rho_{\rm n}
\nonumber\\
& = & \frac{m_{\rm{pl}}}{2\sqrt{n}} c \cdot
\frac{2G}{c^2}\sqrt{n}m_{\rm{pl}} =
\frac{G}{c}\cdot\frac{c\hbar}{2G}=\frac{\hbar}{2}
\nonumber\\
\Rightarrow s_{\rm{z}}  & = &  \pm \,  \frac{\hbar}{2}
\label{spin}
\end{eqnarray}
The spin of a stimulated black hole particle is by this of half
Planck's quantum which
corresponds to a fermion\footnote{However the value of the spin
$ \mid \vec{L} \mid = \frac{G}{c} m_{\rm{pl}}^2 $
relates to the definition of the Planck mass:
If one attaches instead of one two Compton wavelengths
to one Schwarz\-schild-radius
one gets $\sqrt{\frac{c\hbar}{G}}$ for the
Planckmass and a Spin of $\hbar$,
which corresponds to a boson.}.
Particles relating to the energy of stimulation of 
a virtual-miniature-black-hole (\ref{dmn}) will be herein
referred to as SBH-particles.

\section{Discussion}

\begin{table*}

\caption{Chart of SBH-particles:
The entries in italics for the  'SBH-Massless'
$(n\rightarrow\infty)$ and the
'SBH-Universe' $(n \rightarrow 0)$ are values introduced by an
extrapolation of the formulas
of chapter \ref{stNSL} to their limits. The value for the
SBH-massless arises if one gives the mass of the Universe
from  (\ref{MU}) for the virtual mass $m_{\rm n}$ , and the
value of the mass of the SBH-Universe arises if one gives the
mass of the Universe for the stimulated mass $ \Delta m_{\rm n}$.
The mass of a quark is herein estimated as $\sim 10$ MeV. }

\label{tab}
\[\begin{tabular}{|l|l|ll|l|l|l|}
\hline
Eigenvalue & virtual mass & stim. mass &
& SS-diam. & Hawk. time & Name
\\
$ n $ & $m_{\rm{n}}$  & $\Delta m_{\rm{n}} $  & $\Delta E_{\rm{n}} $
& $2\rho_{\rm{n}}=\lambda_{\rm{C}}$ & $\sim t_{\rm{V}} $
& of stim. mass
\\
$ [1] $ & $[\rm{kg}]$ & $[\rm{kg}]$ & $[\rm{eV}]$ & $ [\rm{m}] $
& $ [\rm{sec}] $  & $m_{\rm 0}=\Delta m_{\rm n}$
\\\hline &&&&&&\\
$  \sim 0 $ & $ \pm 8.3\cdot10^{-70} $ & $ 1.4\cdot10^{53} $
& $7.8\cdot10^{88}$
& $  2.5\cdot10^{-96} $ & $ \approx 0
$ & $ \textit{SBH-Universe} $
\\
$  1 $ & $ \pm 1.5\cdot10^{-8} $ & $ 7.7\cdot10^{-9}$
& $0.4\cdot10^{28}$
& $ 4.6\cdot10^{-35} $ & $ \sim10^{-43} $ & Planck quant
\\ &&&&&&
\\
$  2.1\cdot10^{37} $ & $ \pm 7.1\cdot10^{10} $ & $ 1.7\cdot10^{-27}$
& $938\,\rm{[MeV]}$
& $ 2.1\cdot10^{-16} $ & $ \sim10^{12} $ &  SBH-proton
\\
$ 1.9\cdot10^{41} $ & $ \pm 6.6\cdot10^{12} $ & $  1.8\cdot10^{-29}$
& $10\,\rm{[MeV]}$
& $ 2.0\cdot10^{-14} $ & $ \sim10^{18} $ &  SBH-quark
\\
$ 7.2\cdot10^{43} $ & $ \pm 1.3\cdot10^{14} $ & $ 9.1\cdot10^{-31}$
& $0.51\,\rm{[MeV]}$
& $ 3.9\cdot10^{-13} $ & $ \sim10^{22} $ &  SBH-elektron
\\ &&&&&&
\\
$ 8.7\cdot10^{121} $ & $ \pm 1.4\cdot10^{53} $ & $ 8.3\cdot10^{-70}$
& $0.46\cdot10^{-33}$
& $ 4.3\cdot10^{26} $ & $  \sim10^{139} $ &
$ \textit{SBH-massless} $
\\
\hline
\end{tabular}
\]
\end{table*}

\subsection{SBH-Particles}

Table  (\ref{tab}) shows the values of some selected masses referring
to the formulas of chapter \ref{stNSL} and gives an idea of an
Universe which
is build by a fractal manner of black-hole-topologies.
All masses are initiated by stimulated curvatures of space-time
where every black hole has its
typical quantum. For macroscopic black holes these quantums have
energies much smaller than
the electron rest mass in the area of (practical) massless
\footnote{Also in classical quantum 
dynamics a massless particle may have at least a very small
rest mass $\approx 0$ according to
the uncertainty relation$
\Delta E \Delta t = \Delta E \cdot t_{\rm 0} \ge \hbar/2
\Leftrightarrow
\Delta E \ge \hbar/(2t_{\rm 0})
= \frac{3}{4} \hbar H_{\rm 0} \approx 10^{-33} \rm{eV}
$ if one considers a particle blurred over the whole Universe.}
particles like
neutrinos or photons: a macroskopic black hole seems to radiate
thermal like a black body.
But for virtual-miniature-black-holes the stimulated energies are
in the typical range of the
well known  elementary particle restmasses matching their typical
properties as rest energy, Compton
wavelength and spin. 

The Hawking\footnote{
S. Hawking proposed  'orderly' primordial black holes in the mass-range
in question of  $10^{11}$
to $10^{15}$ kg, which should explode as an effect of the
Hawking-radiation just nowadays,
but couldn't be observed yet.
} times $t_{\rm V}$ are a rough hint for the lifetimes of
SBH-particles which show
meaningful values  for stable particles in the range of
typical masses for elementary particles.
But effects 
of quantum gravitation should generate very different values
for this lifetimes, namely those
experimental seen values which can be much less for instable
and much more for stable particles.

The product of virtual mass and stimulated mass is always a
constant for every SBH-particle:
\begin{eqnarray}
C^{\pm}_{\rm{m_{\rm n}}} &:=  \Delta m_{\rm n} \cdot m^{\pm}_{\rm n}
= \pm \frac{1}{2} m_{\rm{pl}}^2
= \pm \frac{c\hbar}{4G}        \nonumber
\\ &=\pm 1.185\cdot10^{-16} \, \rm{kg}^2
\label{Cmn}
\end{eqnarray}
So large black holes have small stimulated masses and small
black holes have large stimulated
masses.
The basic eigenvalue $n=1$ of self-curvature is given by the
Planck mass which has a stimulated
mass of approximately the same quantity.
The sizes of the self-curvatures of space-time increase with $n$
and reach their maximum at
$\sqrt{n}= 9.3\cdot10^{60}$ with the Universe as the largest
Schwarz\-schild area.
An extrapolation of the $\Delta m_{\rm n}/m_{\rm n}$ dependence for an
elementary SBH-particle to the (not allowed) eigenvalue $n=0$ gives
cause to a speculation of
a nearly massless particle of which the stimulated mass is an
Universe with an incredible small
lifetime that satisfies the uncertainty relation.

Through the extrapolation of this supposition the stimulated mass
of an SBH-Universe is a massless particle (like neutrino or photon)
with a rather infinite lifetime, 
and the stimulated mass of a massless particle is an Universe with
a rather incredible short lifetime.
As follows from chapter \ref{stNSL} these quantums may be fermions.
The most light-weight is a
neutrino and the most heaviest a Planck quantum. The speculative
extrapolation to the
(not allowed) eigenvalue $n=0$ gives as the heaviest fermion an Universe.
Here one may imagine a possible
fractal construction of the world build up by black-areas of different sizes.
The observation problem of the
Kopenhagen interpretation of quantum dynamics
~\cite{davies}, which means 'who watches the Universe',
could be brought closer to an explanation
through the assumption that the Universe watches itself or the
Universes watches themselfs:  'The
snake is eating its own tail'
(S.Glashow in Luminet~\cite{luminet}).

\subsection{The entropy of black holes} \label{entropy}

A fundamental question in gravity physics is to explain the high entropy of
black holes. The Bekenstein-Hawking-entropy~\cite{horowitz} is
$S_{\rm{BH}}=\frac{A}{4G\hbar}$. From the above follows by
setting in the surface area
of the black hole $A=4\pi\rho_{\rm{n}}^2$:
\begin{eqnarray}
S_{\rm{BH}}=\frac{2\pi}{c^3}\cdot n
\end{eqnarray}
which relates the black hole entropy directly to its excitation level $n$.
The herein proposed black area for elementary particles should therefore
be a black membrane as proposed by string theory~\cite{horowitz}. Its size
should be of the dimension like it is related to elementary particles
by the Compton/deBroglie-relation and the uncertainity-relation.

If one relates the rest-energy of a particle, using the herein derived
mass-quantization
(\ref{mn})(\ref{dmn}), to the Hawking-temperatur of a Black-Hole, which is 
$T_{\rm{H}}=\frac{\hbar c^3}{8\pi k_{\rm{B}}G M }$ , one gets:
\begin{equation}
\frac{E_{\rm{0}}}{T_{\rm{H}}}
=\frac{\Delta m_{\rm{n}} c^2}{T_{\rm{H}}(m_{\rm{n}})} = 2\pi k_{\rm{B}}
\end{equation} 
This shows that the SBH-particle is in thermodynamical
equilibrium\cite{kiefer} with the Hawking-
temperature of the stimulated-black-hole.

\subsection{The baryogenesis} \label{baryogenese}

The present model takes the energy to create elementary particles
from the energy of
gravitational waves. As a source for this gravitational energy
one may bring in the Universe of
the big bang or big bounce.
The gravitational radiation power~\cite{sexl} of two critical
objects with
dimensions compareable to their Schwarz\-schild radii is:
\begin{equation}
P \approx \frac{c^5}{G} \cdot
\left( \frac{R_{\rm 1}}{r} \right)^3  \cdot
\left( \frac{R_{\rm 2}}{r} \right)^2
\label{PSL}
\end{equation}
in which $r$ is the distance of the objects and $R_1$ and $R_2$
the Schwarz\-schild radii of the objects. This approximation is valid if
$R_{\rm 1}=R_{\rm 2}=r$ as for a collapsing black hole.
Therefore the estimated gravitational radiation power of a collapsing
black hole is
$P \approx \frac{c^5}{G}$.
This power is independent on the mass of the black hole and will be
provided by any black hole as
far as $R_{\rm 1}\approx R_{\rm 2}\approx r$
is valid. For a black-hole-Universe this time is about $t_{\rm 0}$.

The rest energy of the Universe at all is $E_{\rm U} = M_{\rm U}c^2$
which is converted in the time $t_{\rm 0}$ since the big bang.
The average power over all is with  (\ref{cMU}):
\begin{equation}
\langle P_{\rm U}\rangle \cong \frac{E_{\rm U}}{t_{\rm U}}
= \frac{M_{\rm U}c^2}{t_{\rm 0}}
=  \frac{c^3}{H_{\rm 0}G} c^2 \cdot \frac{3H_{\rm 0}}{2}
= \frac{3}{2}  \frac{c^5}{G} > \frac{c^5}{G}
\end{equation}
This means that the gravitational power of a black-hole-Universe
at all should be large enough to
create all the mass it contains.

The energy density of the  $ \Delta m_{\rm n} $ and $m_{\rm n}$
related to the Schwarz\-schild
volume of the related virtual mass $m_{\rm n}$ is:
\begin{eqnarray}
\varepsilon_{\rm{\Delta}} & = & \frac{\Delta E_{\rm{n}}}{\rm{Vol.}}
\cong \frac{\Delta m_{\rm n}c^2}{\frac{4}{3} \pi \rho_{\rm n}^3}
= \frac{3c^7}{32\pi G^2 \hbar} \cdot \frac{1}{n^2}
\nonumber\\
\varepsilon_{\rm{n}}  & = & \frac{E_{\rm{n}}}{\rm{Vol.}}
\cong \frac{m_{\rm n}c^2}{\frac{4}{3} \pi \rho_{\rm n}^3}
= \frac{3c^7}{16\pi G^2 \hbar} \cdot \frac{1}{n}
\label{epsilon}
\end{eqnarray}

The gravitational power  $P$ falls off with the fifth power of the
dimension $r$ of the object as
(\ref{PSL}) shows and close to the singularity  $P(t)$ diverges.
So the following integration is
made for a black-hole-Universe starting at the Planck dimension
with $ t_{\rm{pl}} =
\frac{\hbar}{m_{\rm{pl}}c^2} = 0.76 \cdot10^{-43}\,\rm{sec}$.
The time-dependent evolution of the gravitational power can be
approached with this as follows:
\begin{equation}
\int_{\rm{0}}^{\infty}P(t)dt \cong
\int_{  t_{\rm{pl}}  }^{  t_{\rm 0}  } \frac{\alpha}{r^5(t)}
\frac{c^5}{G}  dt
= M_{\rm U}c^2
\end{equation}
Inserting for $r(t)$ the expansion of the EdS model as referred
in (\ref{EdSSkala}) the constant
$\alpha$ can be determined and one gets after some
elementary calculations:
\begin{equation}
P(t)= (\frac{2^{4/3}\cdot7}{3})\cdot \left(
\frac{G\hbar^7}{H_{\rm 0}^6c^5} \right)^{1/6} \cdot
\frac{1}{t^{10/3}}
\label{PvT}
\end{equation}
From this the time $t_{\rm B}$ of baryogenesis through gravitational
waves can be calculated.
This is the time in which the energy density of $\varepsilon_{\rm{n}}$
(\ref{epsilon}) for a
virtual Schwarz\-schild area  was available.
\begin{equation}
\varepsilon_{\rm{n}}=\frac{E_{\rm{n}}}{\textrm{Vol.}}
\cong
\frac{P(t_{\rm B})\cdot t_{\rm n}}{ \frac{4}{3} \pi \rho_{\rm n}^3 }
= \frac{P(t_{\rm B})\cdot 3c^2}{8 \pi n G \hbar}
\end{equation}
in which $t_{\rm n}$ is the time for traversing $\rho_{\rm{SS}}(m_{\rm n})$.
Inserting (\ref{PvT}) and solving for $t_{\rm B}$ gives:
\begin{eqnarray}
t_{\rm B} &=& (\frac{28\cdot2^{1/3}}{3})^{3/10}
\Bigl[ \left( \frac{G\hbar}{c^5}  \right)^7
\cdot \frac{1}{H_{\rm 0}^6} \Bigr]^{\frac{1}{20}}  \nonumber  \\
&=& 1.643\cdot \frac{t_{\rm{pl}}^{0.7} }{ H_{\rm 0}^{0.3}}
\cong 2\cdot10^{-25}\,\rm{sec}
\label{tB}
\end{eqnarray}
The time
$t_{\rm B}$
is independent on the eigenvalue $n$ and only has a small
dependence on the value of the hubble
constant because the
$-3/10$-power of $H_{\rm 0}$ doesn't matter so much
as the $7/10$-power of $t_{\rm{pl}}$.
Great standardization theories~\cite{coughlan} (GUT)
predict the baryogenesis for the time
between $10^{-35}\, \rm{sec}$ and $10^{-10}\, \rm{sec}$ after the big bang.
The time value for a baryogenesis through primordial
gravitational waves  (\ref{tB})
is compatible with this prediction.

 \subsection{The mass of the proton}

As (\ref{tB}) points out the gravitational power of the evolving Universe
is greater than the power
needed to create virtual black holes for the first primordial
$t_{\rm{B}}\approx2\cdot10^{-25}\,\rm{sec}$.
This time may be interpreted as a kind of phase change as the
Universe stops boiling.
On the other hand
the uncertainity equation for the Universe at this time gives a
mass-equivalent of $m_{l}=\frac{\hbar}{2c^2t_B}$.
Inserting (\ref{tB}) gives:
\begin{equation}
m_l=0.239\cdot(\frac{\hbar^{13}H_0^6}{c^5G^7})^{\frac{1}{20}}
\cong1.8\cdot m_p
\end{equation}
for a $H_{\rm{0}}=87 \frac{\rm{km}}{\rm{Mpc}\cdot\rm{sec}}$,
which relates close
to the proton mass as a limiting upper value for SBH-masses.

%Also the massfunktion
%$m(t_{\rm{B}})=\frac{1}{2}\frac{c^3}{G}\cdot t_B$
%(\ref{kleinm}) shows the same limit if one considers the value of
%the related stimulated mass $\Delta m$ (\ref{dmn1}):
%\begin{equation}
%\Delta m(t_{\rm{B}}) = \frac{1}{2}\frac{m_{\rm{pl}}^2}{m(t_{\rm{B}})}
%=\frac{\hbar}{2c^2t_B}
%\end{equation}

 \subsection{The mass of the neutrino}

In this model, the neutrino is the less weighted fermion, resulting
from the fermionic excitation
of the Universe as the underlying virtual-mass-particle.
The mass relation (\ref{dmn1})
$\Delta m_{\rm n} = \frac{1}{2} \cdot \frac{m_{\rm{pl}}^2}{m_{\rm n}}$
gives a minimum mass for
the lightest neutrino of
$m_{\nu}\geq
\frac{\hbar H_{\rm{0}} }{4c^2}
=0.46\cdot10^{-33}\rm{eV}$,
which can also be interpreted as
the minimum mass for a particle, blurred over the whole Universe,
like shown by Heisenbergs
uncertainity relation. The relation (\ref{dmn1}) shows resemblance
to the Dirac-mass-
term\cite{senjanovic}, which follows from the socalled see-saw-mechanism
in GUT-theory:
\begin{equation}
m_{\rm\nu} \cong \frac{m_{\rm{D}}^2}{M_{\rm R}}
\nonumber
\end{equation}
In this case, the Dirac-mass $M_{\rm{D}}$ could be identified
with the Planckmass $m_{\rm{pl}}$
and the mass of the right-handed neutrino $M_{\rm{R}}$ with the
mass of the Universe.

 \subsection{The photon and the mass of the Universe}

From the theory of nucleosynthesis, which uses the relation of
the occurence of
$^2\rm{D},^3\rm{He},^4\rm{He},^7\rm{Li}$, a fraction of photons
to baryons was
derived\cite{coughlan}:
\begin{equation}
\eta=\frac{n_{\rm{B}}}{n_{\rm{\gamma}}}=(4\pm1)\cdot 10^{-10}
\end{equation}
From the measured photon-density, which is about $n_{\rm{\gamma}}=400/cm^3$,
the mass density of the Universe, mainly protons, should be  
about 10 to 12 percent of the critical mass density\cite{coughlan} 
$0.1\cdot \rho_{\rm{c}}\leq\rho_{\rm{B}}\leq 0.12\cdot \rho_{\rm{c}}$. 

As most of these photons are cold ones, having the temperature
of the cosmic-background-radiation
2.7 K, photons give not a worth mentioning amount of energy to the
mass density of the Universe.
But these photons are 'late' photons, which means that these
photons origin somewhere else in the
Universe and are much red-shifted due to the expansion.
In the SBH-model, the photon can be identified with a bosonic
excitation of the Universe (i.e.
space-time). As the Universe is isotropic and homogenous at every point,
one has to estimate the
energy of the photon as a typical photon like it is emitted by the
photospheres of stars in our
neighbourhood, which have much higher temperatures or typical
wavelenghts around $500$ nm. With
this one gets a fraction of $E_{\rm{\gamma}}=hc/\lambda$ as
\begin{equation}
\frac{E_{\rm{\gamma}}/\eta}{m_{\rm{p}}c^2}
=\frac{h}{c\lambda\eta m_{\rm{p}}}=6.607
\end{equation}
which is more than 6 times the energy-density of the baryonic mass density.
For this example,
with a baryonic mass density of $\rho_{\rm{B}}= 0.131\rho_{\rm{c}}$ and
a averge photon
wavelength of $\lambda_{\rm{\gamma}}=500 \rm{nm}$ at its origin,
the photon would bring up the
missing 86.9 percent of the critical Universe-mass-density
\footnote{At the beginning the Universe was indeed radiation dominated:
the photon energy-density of the
early Universe was a multiple of the baryonic density and this
situation continued for about 1
Million years.~As the Universe expanded, the wavelength of the photon
expanded as well, and the
energy of the photon seems to vanish. In the SBH-picture, only the
energy-density of the
electromagnetic waves decreases, but not the overall energy.
See also page 128 ff. of the book of
R.u.H. Sexl\cite{sexl}}.
In this interpretation, the photon energy could give a considerable
contribution to the mass
density of the Universe: the average-photon-energy and also a not
vanishing mass of the neutrino
could close the Universe without the need of exotic particles.

 \subsection{The electromagnetic force}

Moreover a black hole has not to gravitate because as an effect
of its event-horizon no
gravitational waves or gravitons may leave it. An ordinary black hole
gravitates because it
leaves behind the gravitational field of a collapsing object.
If e.g. a Daemon would cut out a stellar black hole exactly at the
Schwarzschild-border and place
it somewhere else in the Universe, such a  black hole would not
gravitate except by the small
mass equivalent of its Hawking radiation. This mechanism defines
elementary particles as the Hawking
radiation of virtual-miniature-black-holes.

The main difference to classical quantumdynamics is the assumption
that the Compton-wavelength of a particle is assoziated with a
virtual black hole or a black brane, maybe a D-brane like suggested
in string-theory,
which means a quantum space-time curvature of
this diameter. 
Such a virtual-miniature-black-hole is charged with the
constant $C^{\pm}_{\rm{m_{\rm n}}}$ (\ref{Cmn}). This charge is independent
on the mass of the virtual black hole and could give rise to an
electrical charge by quantum gravity effects: The gravitational force
between a non-gravitating virtual-black-hole and a gravitating
stimulated-mass would be $2n$-times stronger than the gravitational force
between two stimulated masses.

The classical ratio between the gravitational force
$F_{\rm{G}}=G\frac{m^2_{e}}{r^2}$ and the
electromagnetic force $F_{\rm{Q}}
=\frac{1}{4\pi\varepsilon_{\rm{0}}}\cdot\frac{e^2}{r^2}$ for an
electron in any given distance $r$ is:
\begin{equation}
\frac{F_{\rm{Q}}}{F_{\rm{G}}}
=\frac{e^2}{4\pi \varepsilon_{\rm{0}} G m^2_{\rm{e}} }
=4.167\cdot 10^{42}
\end{equation}
This ratio lies just between the eigenvalues for SBH-quarks
and SBH-electrons.
If the virtual mass of a SBH-particle would be present as a
higher order effect
in any unknown way, the gravitational force between a
SBH-particle and the virtual-SBH-mass of
its vis-a-vis would be: 
\begin{equation}
F_{\rm{G-virtual}}=G\frac{\Delta m_{\rm{n}}\cdot m_{\rm{n}}}{r^2}
= \frac{c\hbar}{4r^2}
\end{equation}
As this force is independent on $n$ and therefore independent on the
mass of the particle
one may relate it to an electromagnetic charge:
\begin{equation}
F_{\rm{Q}}=\frac{1}{4\pi\varepsilon_{\rm{0}}}\cdot\frac{Q^2}{r^2}
=F_{\rm{G-virtual}}=
\frac{c\hbar}{4r^2}
\end{equation}
which gives
\begin{equation}
Q=\pm \sqrt{\pi\varepsilon_{\rm{0}}c\hbar}=\pm 5.853\cdot e
\end{equation}
and is about 6 times the elementary charge. As this assumption is
just a plain one (as it must be an effect of higher order), this
charge is not so far away from unity as it seems at first sight.
Maxwells equations of
electrodynamics should be an outcome of a theory of quantum
gravitation and for this reason
should give some hints to the formulation of quantum gravity.

 \subsection{Interactions of SBH-particles}

A crucial requirement for SBH-particles is that this particles
should behave like known
particles. So what happens if two SBH-particles collide ?
In this plain model, SBH-particles have at least
two quantum-numbers: The spin $s_{\rm{z}}=\pm \hbar/2$ and the
virtual-mass-charge
$C^{\pm}_{\rm{m_{\rm n}}}=\pm c\hbar/4G$, which corresponds to the
electrical charge $\pm e$ and
a virtual-mass of $\pm m_{\rm n}$. 

A SBH-positron has the opposite values of this
 quantum-numbers, so SBH-electron and SBH-positron annihilate to a
radiation of 2-times the energy of the stimulated mass $\Delta m_{\rm n}$,
just like electron and positron do.
When two SBH-electrons meet, they will not merge to a
double-massive SBH-electron, as the two
$\hbar/2$-spins would add to $\hbar$ or $0$,
but the double-massive SBH-electron would have a $\pm \hbar/2$-spin.

Recent experiments\cite{daviss} in a quantum-Hall environment
show the possibility of the creation
of fractional charged pseudo-electrons. This splitted electrons, with e.g.
$e/3$ or $e/5$ charges, are created between groups of
close circulating Hall-electrons.
They seem to behave like one would assume for SBH-electrons: in their close
vicinity space-time can be bend enough to form a virtual fractional-electron.

In the SBH-picture, the photon is the most elementary particle:
Massive elementary particles are
photons 'captured' by virtual-black-holes. And indeed, every
elementary particle can be
transformed to photons by 
annihilation~\footnote{Another process to do this, is the capture
of a particle by an ordinary black hole. The
captured particle adds its mass to the black hole and after a while it
is radiated away by
Hawking-radiation and the black hole restores its mass as it was
before the capture. By this
mechanism, also known as the information-paradoxon, the particle is
transformed to a photon
loosing its quantum numbers.}.

\subsection{Resume}

A simple model of the Universe and its elementary parts was derived
by assuming
that a gravitational wave should not overrun its own event-horizon.
From this assumption follows the relation between masses, Compton-size,
and spins
of fermions and a vivid interpretation of the
particle/wave-dualism is given.
The origin of electromagnetic charge as an effect of the virtual
mass of a particle was proposed.
Also follows the quantization of black holes, as was also shown e.g.
by Khriplovich\cite{khriplovich} by dimensional arguments.

All matter should be build up by the event-horizons of gravitational waves.
Every natural wave, like waves running with the speed of light or sound,
are kinds of black-branes: No information will leave the wavefront
to the outer regions.
As gravitation gravitates, these waves build up closed and stable regions,
defining fermionic elementary particles and even the Universe itself.
The Hawking-radiation of the black-branes of the derived particles resemble
the masses of the known fermions.

\end{document}